\documentclass[amsmath,twocolumn,aps, longbibliography]{revtex4-1} %nofootinbib
\usepackage{amsthm,amsfonts,graphicx,verbatim, color}
\usepackage[french]{babel}
\usepackage{bm}% bold math
\usepackage[utf8]{inputenc}
\let\oldmarginpar\marginpar
\renewcommand\marginpar[1]{\-\oldmarginpar[\raggedleft\footnotesize #1]%
{\raggedright\footnotesize #1}}

\newcommand{\be}{\begin{equation}}
\newcommand{\ee}{\end{equation}}
\newcommand{\bea}{\begin{eqnarray}}
\newcommand{\eea}{\end{eqnarray}}

\renewcommand{\epsilon}{\varepsilon}

\usepackage[normalem]{ulem}

\def\beq{\begin{equation}}
\def\eeq{\end{equation}}
\def\bea{\begin{eqnarray}}
\def\eea{\end{eqnarray}}

\newcommand{\affuzh}{\affiliation{Department of Physics, University of Zurich, Winterthurerstrasse 190, 8057 Zurich, Switzerland}}

\begin{document}

\title{Numerical investigation of gapped edge states in fractional quantum Hall-superconductor heterostructures}
\author{C\'ecile Repellin}
\affiliation{Max-Planck-Institut f\"ur Physik komplexer Systeme, 01187 Dresden, Germany}
\author{Ashley M.\ Cook}\affuzh

\author{Titus Neupert}\affuzh
\author{Nicolas Regnault}
\affiliation{
Laboratoire Pierre Aigrain, D\'epartement de physique de l'ENS, \'Ecole normale sup\'erieure, PSL Research University, Universit\'e Paris Diderot, Sorbonne Paris Cit\'e, Sorbonne  Universit\'es, UPMC Univ. Paris 06, CNRS, 75005 Paris, France
}

\begin{abstract}
Fractional quantum Hall-superconductor heterostructures may provide a platform towards non-abelian topological modes beyond Majoranas. However their quantitative theoretical study remains extremely challenging.
We propose and implement a numerical setup for studying edge states of fractional quantum Hall droplets with a superconducting instability. The fully gapped edges carry a topological degree of freedom that can encode quantum information protected against local perturbations. We simulate such a system numerically using exact diagonalization by restricting the calculation to the quasihole-subspace of a (time-reversal symmetric) bilayer fractional quantum Hall system of Laughlin $\nu=1/3$ states. We show that the edge ground states are permuted by spin-dependent flux insertion and demonstrate their fractional $6\pi$ Josephson effect, evidencing their topological nature and the Cooper pairing of fractionalized quasiparticles. 
% This offers the first quantitative (/microscopic?) verification of hypotheses central to the prediction of non-abelian modes at the edge of fractional quantum Hall systems.
The versatility and efficiency of our setup make it a well suited method to tackle wider questions of edge phases and phase transitions in fractional quantum Hall systems.
\end{abstract}
\maketitle

\section{Introduction}\label{sec:Introduction}
The fractional quantum Hall~\cite{Tsui-PhysRevLett.48.1559} (FQH) effect harbors a variety of exotic topologically ordered quantum phases, from the well understood Laughlin states all the way to states~\cite{Moore1991362, Read-PhysRevB.59.8084} with non-Abelian quasiparticles such as Majorana fermions and Fibonacci anyons. The experimental exploration of these systems faces two challenges: First, the desired more exotic topological orders, which could be used, e.g., for universal quantum computation~\cite{kitaev-2003AnPhy.303....2K}, can only be accessed under extreme experimental conditions, as they are protected by very small energy gaps. Thus, despite intense efforts, definite experimental confirmation of the non-Abelian nature of a FQH phase is still lacking. Second, the topological information is encoded in degenerate ground states or the state of quasiparticles, and is therefore intrinsically hard to measure and manipulate.

To overcome both of these obstacles, several recent studies proposed focusing on more conventional FQH states, such as the Laughlin $\nu=1/3$ state, and to add twist defects~\cite{Bombin-PhysRevLett.105.030403,Barkeshli-PhysRevB.88.241103,Barkeshli-PhysRevB.87.045130,Barkeshli-2014arXiv1410.4540B}. Among the various physical implementations of these twist defects, one deliberately couples the edge states~\cite{Clarke-2013NatCo...4E1348C,Lindner-PhysRevX.2.041002,Cheng-PhysRevB.86.195126,Cong-PhysRevLett.119.170504,Meidan-PhysRevB.95.205104, Klinovaja-PhysRevB.90.155447}. In these proposals, the topological excitations could be localized at domain walls of differently gapped edge segments. Advantages of this approach include the comparably large gap of the Laughlin $\nu=1/3$ state, which protects the quantum state, and the localized nature of the topological excitations, which facilitates their measurement and manipulation. For example, Refs.~\onlinecite{Mong-PhysRevX.4.011036,Vaezi-PhysRevX.4.031009} use parafermion excitations at domain walls between magnetic and superconducting gapped regions of FQH edges to build topological order akin to the $\mathbb{Z}_3$ Read-Rezayi state. Barkeshli subsequently pointed out that topological information is also stored in a pair of counter-propagating $\nu=1/3$ edge states that are fully gapped out by a superconducting order parameter~\cite{Barkeshli-PhysRevLett.117.096803}. The gapped edge, on which a pairing between fractionalized quasiparticles is induced, has a well defined quantized total charge that can take the values $0$, $2e/3$ and $4e/3$ (modulo the Cooper pair charge $2e$). This nonlocal observable defined along the (closed) edge distinguishes three topologically degenerate ground states of the edge. By appropriately coupling several gapped edges, one can in principle manipulate their topological ground state~\cite{Barkeshli-PhysRevLett.117.096803}. Another approach to engineer parafermion excitations from Abelian topological order relies on lattice defects and was recently implemented numerically in Refs.~\onlinecite{Liu-PhysRevLett.119.106801,Vaezi-2017arXiv170601192V}.
The FQH edge states are also a convenient system to study the bulk-boundary correspondence in topologically ordered systems. Unlike noninteracting symmetry protected phases in two spatial dimensions, interacting integer and fractional quantum Hall states can support several distinct edge phases with different universal properties but the same symmetries~\cite{Cano-PhysRevB.89.115116, Barkeshli-2015arXiv150706305B}.

\begin{figure}[h!]
\begin{center}
\includegraphics[width=0.45\textwidth]{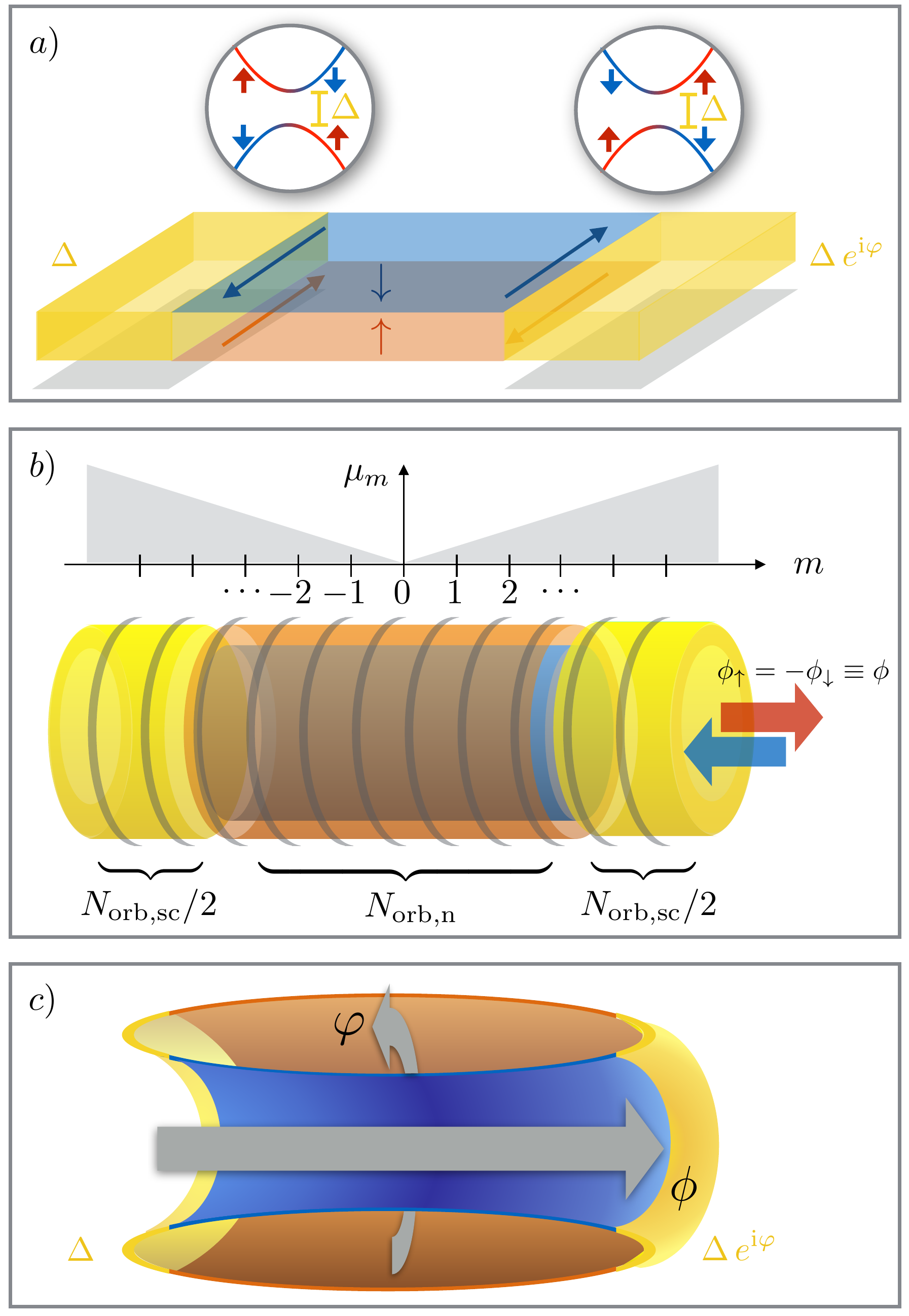}
\caption{
Schematics of the physical geometry and the one used for the numerical investigation.
a) Fractional topological insulator heterostructure in which carriers with spin up and down (red and blue) 
form a fractional quantum Hall state with opposite chirality. Proximity to superconducting reservoirs (yellow) induces a superconducting gap in their edge channels. To study the Josephson effect, relative phase $\varphi$ between the left and right superconducting order parameter is included.
b)
When imposing periodic boundary conditions along the edges, resulting in a cylinder geometry, each edge carries a  topological degree of freedom. The boundary conditions can be twisted by inserting a flux $\phi$ into the cylinder for spin up electrons and $-\phi$ for spin down electrons. In the Landau gauge orbitals are localized along the cylinder, where we consider $N_{\mathrm{orb,n}}$ and $N_{\mathrm{orb,sc}}$ normal and superconducting orbitals, respectively. 
 The typical separation between orbitals is $\frac{2\pi \ell_B^2}{L_y}$, where $L_y$ is the cylinder perimeter.
The droplet is confined by a linear potential $\mu_m$.
c) 
With the counter-propagating edges gapped out, the bilayer FQH state on the cylinder is topologically equivalent to a single layer FQH state on a torus, where the fluxes $\phi$ and $\varphi$ run through its two noncontractible cycles and can be used to explore its topological ground state degeneracy. It is thus topologically equivalent to the ground state degeneracy of the gapped edge modes.
 } 
\label{fig:CylinderOrbitals}
\end{center}
\end{figure}

While effective models (e.g., using a bosonized description of the edge~\cite{Sagi-PhysRevB.91.245144, Ebisu-PhysRevB.95.075111}) have permitted striking predictions at the edge of topologically ordered systems, open questions remain which can only be addressed by a microscopic approach. First, in the context of the bulk-boundary correspondence, which boundary phase is favored by certain microscopic interactions remains largely unexplored (especially when non-abelian liquids are used as the building blocks).
As for localized edge modes, the braiding of non-abelian excitations relies on the possibility to tunnel quasiparticles through the bulk while keeping the edge gap open\cite{Lindner-PhysRevX.2.041002}.
This hypothesis relies on the hierarchy of energy and length scales in the system. Numerical simulations are necessary to achieve such quantitative analysis, and have the potential to identify challenges that could have been obscured by effective analytical models. They seem indispensable as experiments are undertaking the first steps to realize the ideas outlined above~\cite{Amet-Science2016,Matsuo-2017arXiv170303189M}.

Here, for the first time, we undertake an extensive numerical calculation of a FQH system coupled to a superconductor using exact diagonalization. More explicitly, we consider a bilayer FQH system, with magnetic field perpendicular to the layers, where the orientation of the field for one layer is opposite to that for the other layer. This is equivalent to a time-reversal symmetric fractional topological insulator~\cite{Stern-annurev-conmatphys}. This construction permits gapping out of the edge states with singlet interlayer superconducting pairing. Our calculations are performed on a cylinder geometry in which the bilayer-FQH droplet has two edges. 
% A numerical study of these effects is a priori rather challenging, due to the limited accessible system sizes.
To make numerics feasible, we restrict our study to the subspace of zero energy bulk and edge excitations of the Laughlin $\nu=1/3$ state in each layer. These many-body quantum states being Jack polynomials\cite{Bernevig-PhysRevLett.100.246802,Bernevig-PhysRevLett.101.246806}, we can perform an efficient evaluation of the microscopic model matrix elements in the reduced basis. Our setup can thus be straightforwardly generalized to study the edges of other FQH model states with similar properties but richer topological order (such as the Moore-read state).
It could also be cast into the matrix product state framework~\cite{zaletel-PhysRevB.86.245305,estienne-PhysRevB.87.161112,Zaletel-PhysRevLett.110.236801}, which should provide access to larger system sizes.

The article is organized as follows. We first describe in Sec.~\ref{sec:Hamiltonian} the microscopic model that we consider. We discuss the approximations that allow to perform numerical simulations, especially the projection onto the zero energy subspace of the interaction Hamiltonian. In Sec.~\ref{sec:SpectralEvidence}, we provide the spectral evidence for the gapped edge states, including their threefold degeneracy. Starting from the case without the superconducting coupling, we discuss how the system evolves into the threefold ground state manifold. We then present our study of the charge distribution of the system in Sec.~\ref{sec:ChargeDistribution}. Sections~\ref{sec: spin-dependent flux} and~\ref{sec: Josephson} give the numerical evidence for two key signatures of the  topological degrees of freedom encoded in the gapped edge modes: the ground state manifold mixing under the charge pumping and the $6\pi$-periodic Josephson effect.

\section{Hamiltonian and effective Hilbert space}\label{sec:Hamiltonian}

In the Landau gauge, a single-particle basis that spans the lowest Landau level on a cylinder of circumference $L_y$ is given by
\begin{equation}
\phi_m(x,y)=\frac{1}{L_y \ell_B \sqrt[4]{\pi}} e^{\mathrm{i} \frac{2 m \pi y}{L_y}} e^{-\frac{1}{2\ell_B^2}\left(x-\frac{2\pi m \ell_B^2}{L_y}\right)^2},
\label{CylinderObitals}
\end{equation} 
where $\ell_B$ is the magnetic length which we will set to unity in the rest of the manuscript. We truncate the single-particle Hilbert space of the cylinder by allowing for $m$ to take integer values in the range $-N_\Phi/2\leq m\leq +N_\Phi/2$ for some integer $N_\Phi$. Note that $m$ is either an integer or half integer depending on the parity of $N_\Phi$. Here, $m$ plays the role of both the  $y$-momentum of the wave function and at the same time determines the location of the wave function along the $x$ direction. This coupling of momentum and position is enforced by the lowest Landau level projection.

We now consider a bilayer system, where the two layers are distinguished by a spin index $\uparrow$, $\downarrow$ and the Hall effect in one layer is opposite in chirality to the Hall effect in the other layer. This is the case in so-called fractional topological insulators, where the labels $\uparrow$, $\downarrow$ may correspond to the physical spin and the spin-dependent magnetic field is akin to the spin-orbit coupling. An alternative scenario more relevant to traditional FQH experiments is one in which $\uparrow$, $\downarrow$ label the carriers in two adjacent quantum wells, one being electron-like and the other hole-like. A homogeneous magnetic field gives rise to edge states in both quantum wells and the direction of propagation in one quantum well is opposite to that in the other quantum well.  Our numerical study applies equally well to each of these physical realizations, but we choose to describe our results using the terminology of the fractional topological insulator realization. 
In either case the single particle eigenstates are 
\begin{equation}
\begin{split}
&\psi_m^\uparrow (x,y) = \phi_m(x, y),
\\
&\psi_m^\downarrow (x,y) = \phi^*_{m}(x, y)= \phi_{-m}(-x, y),
\end{split}
\label{eq: single particle eigenstates} 
\end{equation}
with $-N_\Phi/2\leq m\leq +N_\Phi/2$. This ensures that the system is invariant under time-reversal symmetry $\mathcal{T}=K\,\mathrm{i}\sigma_y$ for spinful fermions, where $K$ is complex conjugation and $\sigma_y$ is the second Pauli matrix acting in spin space. Note that none of the topological features we are interested in are protected by $\mathcal{T}$. In fact, in the electron-hole bilayer realization of the system, $\mathcal{T}$ is not the physical symmetry of the system, but an artificial symmetry of the model that may be broken in a microscopic realization.

In the Fock space spanned by the single-particle states $\psi_m^\uparrow$ and $\psi_m^\downarrow$, we consider a Hamiltonian of the form
\begin{equation}
\begin{split}
\hat{H}=&\,\hat{H}_{{\rm 2-body},\uparrow}+\hat{H}_{{\rm 2-body},\downarrow}\\
&+\sum_{m=-s}^{s} \mu_m \Big(\hat{c}_{m,\uparrow}^\dagger \hat{c}_{m,\uparrow} + \hat{c}_{m,\downarrow}^\dagger \hat{c}_{m,\downarrow}\Big) \\
&+\frac{1}{2C} \Big(\hat{N} - N_0 \Big)^2\\
&+\Delta_0 \sum_{m=-s}^{s} f_m \Big(\hat{c}_{m,\uparrow}^\dagger \hat{c}_{m,\downarrow}^\dagger + \mathrm{h.c.}\Big)\,,
\end{split}
\label{eq: general Hamiltonian}
\end{equation}
where $H_{{\rm 2-body},\sigma}$ is the two-body interaction for fermions with spin $\sigma=\uparrow, \downarrow$ and $\hat{c}^\dagger_{m, \sigma}$ creates an electron in state $\psi_m^\sigma$. For $H_{{\rm 2-body},\sigma}$ we use the pseudo-potential Hamiltonian with $V_1$ being the only non-zero pseudo-potential coefficient. The coefficients $\mu_m$ describe a confining potential that is rotationally symmetric around the cylinder axis. The operator $\hat{N}=\hat{N}_\uparrow+\hat{N}_\downarrow$ measures the total number of particles. A schematic representation of this setup is sketched in Fig.~\ref{fig:CylinderOrbitals}.

Due to the presence of mean-field superconductivity, the particle number is not conserved. To tune the system into a regime with finite particle number density, the charging energy of strength $1/(2C)$ has been added to the Hamiltonian. The two parameters $C$ and $N_0$ permit tuning of the average number of particles in the system. Finally, $\Delta_0$ is the overall strength of the superconducting coupling while $f_m$ is the (dimensionless) variation of the superconducting order parameter along the cylinder, assuming a superconducting pairing potential that is rotationally symmetric along the cylinder axis. We will further assume a superconducting pairing potential that is nonzero only at the edge of the FQH droplet. In the electron-hole bilayer realization of the system, the term proportional to $\Delta_0$ takes the role of a charge conserving backscattering term between the layers.

\begin{figure}[t]
\begin{center}
\includegraphics[width=0.45\textwidth]{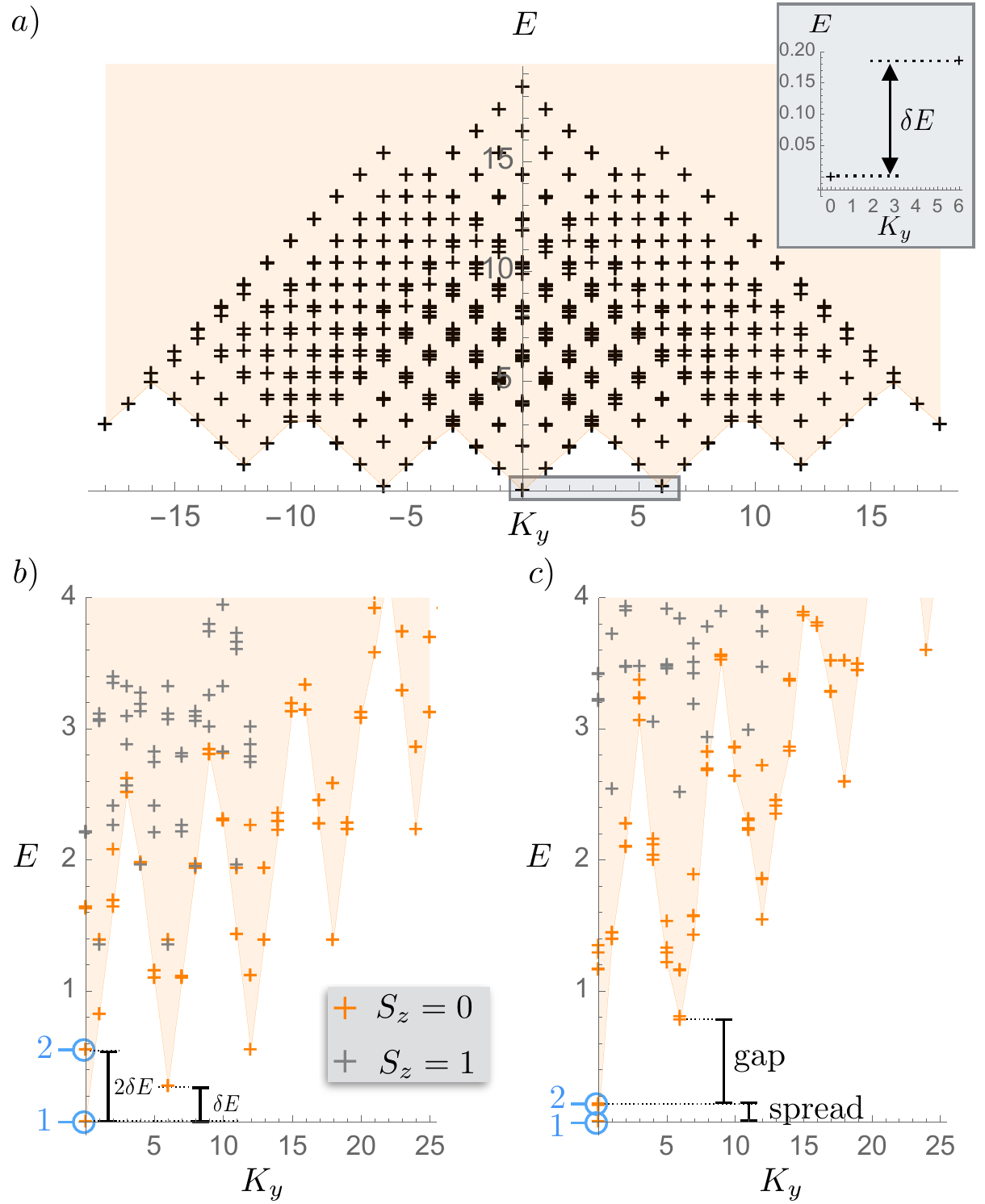}
\caption{
Spectra of Hamiltonian Eq.~\eqref{eq: general Hamiltonian} on a cylinder with 
circumference $L_y/\ell_B=15.0$, $N_\Phi=21$ flux quanta, $1/(2C)=0.256271$, $N_0=30$, and a linear symmetric chemical potential $\mu_m$ as shown in Fig.~\ref{fig:CylinderOrbitals}~b). 
a) Spectrum of a single layer of the cylinder with $6$ fermions. Energies are shifted such that the ground state energy is zero. Shaded areas are a guide for the eye showing the `arcs' in the relative momentum $K_y$ arising from the edge states of the Hall droplet. The inset is a zoom on the low lying state within the gray box, showing the energy difference $\delta E$ between the first excited state and the ground state.
b) and c) are spectra for the full double layer with b) $\Delta_0=0$ while in c) $\Delta_0=2.1$ with $N_{\mathrm{orb,sc}}=16$ superconducting orbitals. Blue figures give the number of states in the circle.  In c) the superconducting gap in the edge states is apparent from the three topological `ground states' moving below the bottom of the next arc. Orange states have total spin $S_z=0$, gray states $S_z=1$, and states with higher spin only appear above this energy window. Both spectra symmetrically extend to $K_y\to -K_y$.
 } 
\label{fig:spectra}
\end{center}
\end{figure}

The Laughlin state edge and bulk quasihole excitations are the exact zero energy states of the model interaction $\hat{H}_{{\rm 2-body}, \sigma}$ at filling $\nu=1/3$. Their corresponding wavefunctions have an analytical expression on the cylinder geometry~\cite{Rezayi-PhysRevB.50.17199}. Being Jack polynomials, they can be conveniently decomposed into the occupation basis~\cite{Bernevig-PhysRevLett.103.206801}. In Ref.~\onlinecite{Soule-PhysRevB.86.115214}, a careful and detailed numerical study of this state and its edge excitations was performed on the cylinder geometry using a confinement similar to the $\mu_m$ term of Eq.~\eqref{eq: general Hamiltonian}. In particular, the low energy spectrum (i.e., below the bulk energy gap) for a finite size quantum Hall droplet has the characteristic shape shown in Fig.~\ref{fig:spectra}~a).

To make progress in the numerical evaluation of Hamiltonian~\eqref{eq: general Hamiltonian}, we send the gap of the FQH state to infinity, i.e., we set $V_1\to\infty$, by projecting the Hamiltonian to the zero-energy subspace of $H_{{\rm 2-body},\uparrow}+H_{{\rm 2-body},\downarrow}$. This is the space of Laughlin quasiholes in each layer. The densest state(s) in this subspace are the Laughlin FQH states with a filling fraction of $1/3$ per layer and spin. With this projection to the Laughlin quasihole space in place, the Hilbert space dimension is dramatically reduced. In a second step, we diagonalize the chemical potential, superconducting, and charging energy terms in this quasihole space.

\begin{figure*}[t]
\begin{center}
\includegraphics[width=0.98\textwidth]{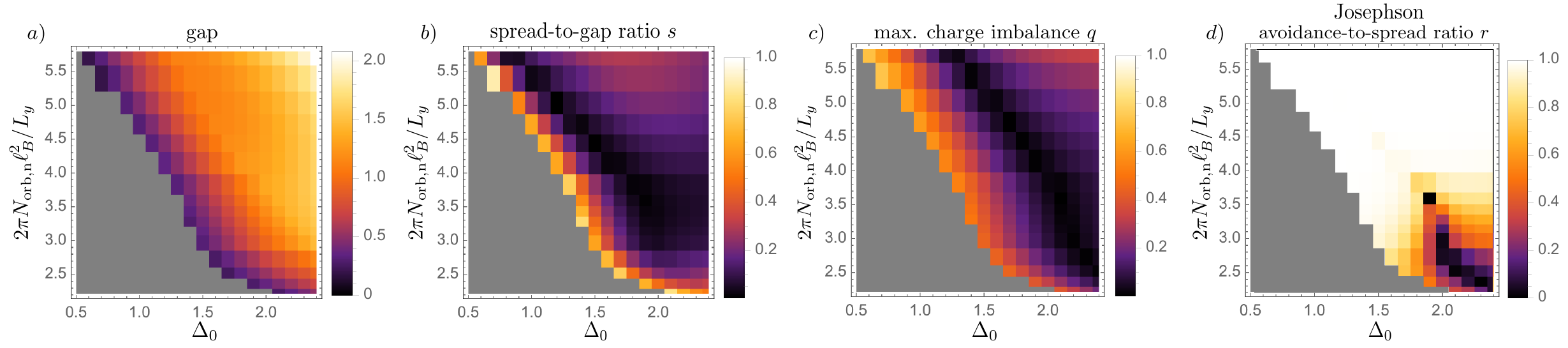}
\caption{
Characterization of the low-energy spectrum of Hamiltonian Eq.~\eqref{eq: general Hamiltonian} as a function of $2\pi N_{\mathrm{orb,n}}\ell_B^2/L_y$ and the strength of the pairing potential $\Delta_0$. $2\pi N_{\mathrm{orb,n}}\ell_B^2/L_y$ approximates the physical distance between the superconducting regions and can be tuned by varying $L_y$. The charging energy is optimized for each $L_y$ using the procedure defined in Supplementary Note II. Other parameters are identical to those of Fig.~\ref{fig:spectra}.
a) Gap between the three lowest states in the $K_y=0$ sector and the next excited states. Gray color indicates that the three lowest states do not have $K_y=0$.
b) Spread of the three lowest states in the $K_y=0$ sector as indicated in Fig.~\ref{fig:spectra}~c) divided by the gap. Gray color indicates the region in which the ratio exceeds 1 or where the three lowest states do not have $K_y=0$.
c) The largest eigenvalue $q$ of the charge imbalance operator defined in Eq.~\ref{eq: delta Q operator} in the space of the tree lowest states (at $K_y=0$). Gray color is used if the three lowest states do not have $K_y=0$.
d) The difference $r$ between the energy of the second lowest eigenstate at Josephson phases $\varphi=0$ and $\varphi=\pi$, normalized by the spread and each time measured with respect to the ground state energy, as defined in Eq.~\ref{eq:RJosephson}. The closer $r$ gets to $1$, the better can the $6\pi$ Josephson effect be observed. The gray color has the same meaning as in c).
 } 
\label{fig:phase diagrams}
\end{center}
\end{figure*}

The parameters $1/2C$ and $N_0$ of the charging energy term are used to control (i) the average number of particles in the droplet and (ii) the energy difference to sectors with nearby particle numbers.
We choose these parameters by diagonalizing the system in the absence of superconductivity. Note that $N_0$ is not equal to the particle number in the non-superconducting ground state, because the $\mu_m$ term also contributes an energy cost that depends on the particle number. We choose $1/2C$ and $N_0$ such that the ground state has a desired particle number $\tilde{N}$ and the lowest energy state in the sector with $\tilde{N}+2$ particles is degenerate with the lowest energy state in the sector with $\tilde{N}-2$ particles. The energy difference between the $\tilde{N}$ and the $\tilde{N}\pm2$ sectors is chosen to be small enough that a moderate superconducting term can couple these sectors (as detailed in Supplementary Note II).

The problem defined by Hamiltonian~\eqref{eq: general Hamiltonian} has two good quantum numbers: these are the total spin measured by $\hat{S}_z=\frac{1}{2}(\hat{N}^\uparrow-\hat{N}^\downarrow)$ (which also encodes the fermion parity) and the relative (angular) momentum $K_y=\sum_{i\in\uparrow}m_i-\sum_{j\in\downarrow}m_j$, where $m_i$ and $m_j$ are the  quantum numbers of the occupied $\uparrow$-spin and $\downarrow$-spin states, respectively, as defined in Eq.~\eqref{eq: single particle eigenstates}. In the electron-hole bilayer realization of the system, $\hat{S}_z$ is the particle number operator.

\section{Results}

\subsection{Spectral evidence for the topological edge states}\label{sec:SpectralEvidence}

We present the spectral features associated with the model defined by Hamiltonian~\eqref{eq: general Hamiltonian} and sketched in Fig.~\ref{fig:CylinderOrbitals}: three low-energy states protected by an energy gap.
We choose a symmetric confining potential $\mu_m=|m|$ that is linear in the orbital space index with a slope of $1$ (which will serve as the unit of energy throughout this paper), as shown in Fig.~\ref{fig:CylinderOrbitals}~b).

\subsubsection{Decoupled layers}

We first study the spectrum of two decoupled layers, i.e., in the absence of superconductivity when $\Delta_0=0$, which is shown in Fig.~\ref{fig:spectra}~b). It can be understood as a combination of two independent spectra of a confined Laughlin $\nu=1/3$ state on a cylinder with a full edge structure as displayed in Fig.~\ref{fig:spectra}~a). 
In the presence of the linear confining potential $\mu_m=|m|$, the Laughlin state on the cylinder has a characteristic spectral feature as a function of $K_y$: focusing on a single layer with $N/2$ particles (assuming $N$ even), the ground state has $K_y=0$. The lowest lying excitations appear in the momentum sector $K_y=\pm N/2$. $\delta E$ denotes the energy difference between these states and the ground state. Further low-lying states are located in the sectors $K_y=n N/2$, $n\in \mathbb{Z}$. The lowest lying states in the other momentum sectors are higher in energy, giving rise to an arc-like structure in the spectrum, as observed in Fig.~\ref{fig:spectra}~a) . These arcs are highlighted by the shaded region in Fig.~\ref{fig:spectra}. The lowest states at momenta $K_y=n N/2$, $n\in \mathbb{Z}$ are those where the droplet has been rigidly moved by $n$ orbitals, giving rise to an extra energy cost of about $n\nu$ due to the higher chemical potential of the now occupied orbitals in comparison to the emptied ones.  Since the center of mass of the wave function is moved by $n$ orbitals, the change in center of mass momentum is $K_y=n N/2$. 
The lowest excitations in other momentum sectors are local edge excitations or combinations thereof. Since they increase the size of the Hall droplet, they cost more energy than the rigidly moved Laughlin state. 

Within these considerations, we can understand the spectrum of Fig.~\ref{fig:spectra}~b) as a finite-size representation of a collection of gapless FQH edge states. In particular we can understand the low-energy structure as superpositions of the states in the two layers $\uparrow$, $\downarrow$. We denote by $|\sigma, 0\rangle$, $|\sigma, \pm N/2 \rangle$ the three lowest states in each of the $\sigma = \uparrow,\downarrow$ sectors which occur at momenta $0$ and $\pm N/2$. 
The state $|\uparrow, 0\rangle\otimes |\downarrow, 0\rangle$ is then the nondegenerate ground state labelled by a blue `1' in Fig.~\ref{fig:spectra}~b). The states $|\uparrow, \pm N/2 \rangle\otimes |\downarrow, 0\rangle$ and $|\uparrow, 0 \rangle\otimes |\downarrow, \pm N/2\rangle$ are four degenerate states at momenta $K_y=\pm N/2$ found at the bottom of the first arc in Fig.~\ref{fig:spectra}~b) at energy $\delta E$ above the ground state. The states  $|\uparrow, N/2 \rangle\otimes |\downarrow, N/2\rangle$ and $|\uparrow, -N/2 \rangle\otimes |\downarrow, - N/2\rangle$ are degenerate at $K_y=0$ and labelled by a blue `2' in Fig.~\ref{fig:spectra}~b). They occur at exactly $2\delta E$ above the ground state. 

\subsubsection{Gapping the three-fold ground state}\label {sec:gapping}

We now compare the low-energy structure of the system with zero [Fig.~\ref{fig:spectra}~b)] and non-zero superconducting pairing in the outer orbitals [Fig.~\ref{fig:spectra}~c)]. The states that were at $2\delta E$ in the former system moved substantially below the ones formerly at $\delta E$. Thus, the spectrum cannot be decomposed into that of two independent layers anymore. Superconductivity coupled the layers. 
Closer inspection also reveals a tiny but nonvanishing lifting of the degeneracy between the two lowest-lying excited states at $K_y=0$. We interpret the three lowest states in the $K_y=0$ sector as the quasi-degenerate topological states of the edges and the gap above them as the superconducting gap induced in the counter-propagating FQH edge modes. 

The three-fold ground state degeneracy of the gapped edge states can be understood as follows.
By introducing a gap, the superconducting coupling turns the bilayer quantum Hall state with edges into a single-layer quantum Hall state on a manifold without boundary. As sketched in Fig.~\ref{fig:CylinderOrbitals}~c), this manifold is topologically equivalent to a torus, where the space between the two layers becomes the interior of the torus. This is in line with the opposite sign of the Hall conductivity in the two layers, because the normal to the torus surface is also reversed. On the torus, a Laughlin state at filling $\nu=1/3$ has a three-fold ground state degeneracy. This degenerate ground state is thus topologically equivalent to the three ground states we observe in the superconducting bilayer system.
Fractional quantum Hall ground states on the torus can be manipulated by inserting flux through the holes of the torus, see Fig.~\ref{fig:CylinderOrbitals}~c). This will be demonstrated in Secs.~\ref{sec: spin-dependent flux} and~\ref{sec: Josephson} via manipulating the flux $\phi$ and the Josephson phase $\varphi$, respectively.

More relevant to the physics of the bilayer heterostructure is an interpretation of the ground state degeneracy in terms of Cooper-paired Laughlin quasiparticles. 
Due to the mean-field superconducting order parameter, the particle number is only defined modulo 2. 
Assuming that the low-energy Laughlin quasiparticles (of charge $e/3$) that comprise the edge mode are Cooper-paired, this leaves three nonequivalent configurations for the charge of one edge: $0$, $2e/3$, and $4e/3$ -- each modulo 2.
Since the total number of particles of the system is quantized to integers, the two superconducting edge states cannot independently support any of these charge configurations. Rather, they either both have charge 0, or the left edge has charge $2e/3$ and the right edge $4e/3$, or vice versa. We thus expect a total of three nearly degenerate topological ground states from this consideration as well, in line with our numerical observation.

Beyond the two special cases shown in  Fig.~\ref{fig:spectra}~b) and~c), we have performed an extensive study of the spectral properties when varying the system parameters. Some results are given in Fig.~\ref{fig:phase diagrams} for the largest system size that can be reached. Another system size is discussed in Supplementary Note I. We fix the total number of superconducting orbitals $N_{\rm orb,sc}$, equally split between the two ends of the cylinder. The selected value is a compromise between fully covering the edge modes and a large enough non-superconducting region of $N_{\rm orb,n}$ consecutive orbitals where an incompressible liquid can develop [as depicted in Fig.~\ref{fig:CylinderOrbitals}~b)]. For each perimeter $L_y$ the charging energy parameter $1/2C$ is optimized as discussed in Supplementary Note II ($N_0$ being fixed for the full diagram to $N_0=30$). Instead of using $L_y$ for the vertical axis, we have plotted the data as a function of the approximate width of the normal region, i.e., $\frac{2\pi \ell_B^2 N_{\rm orb,n}}{L_y}$. Such a quantity is more natural when comparing different system sizes.

In Fig.~\ref{fig:phase diagrams} a), we show the energy gap above the three lowest energy states. We set the gap to zero if these three states do not have $K_y=0$. We also provide $s$, the ratio between the energy spread of the three lowest energy states and the gap as previously defined. We cap $s$ to one or set it to one if the gap is zero or the three lowest energy states do not have the expected quantum numbers. To be able to claim that we have a low energy manifold made of these three states separated by a gap from the higher energy excitations, we need $s<1$. The smaller $s$ is, the closer to an exact degenerate manifold we are. As can be observed in Fig.~\ref{fig:phase diagrams} b), we have a large region where $s$ is small, beyond a critical value of $\Delta_0$ depending on $L_y$.

\subsection{Charge distribution}\label{sec:ChargeDistribution}

In this section we study the charge distribution between the left and right halves of the system. In a physical realization of the system, two scenarios should be distinguished. If the two edges are coupled to the same superconductor, which implies that they are phase coherent, no quantized charge can be associated with one edge alone. In contrast, if the two edges are gapped by independent superconducting reservoirs, they carry independent fractionally quantized charges. However, in this latter case the charging energy is expected to lift the ground state degeneracy and the states are not topological.

The observable that measures the charge disproportionation between the two superconducting edges is given by 
\begin{equation}
\begin{split}
\hat{Q}_{\mathrm{R}}-
\hat{Q}_{\mathrm{L}}
=&\,
\sum_{\sigma=\uparrow, \downarrow}
\sum_m
\int\mathrm{d}x\int_0^{L_y}\mathrm{d}y\,
\mathrm{sgn} (x)\\
&\times
\left|\psi^\sigma_m(x,y)\right|^2 \, \hat{c}^\dagger_{m,\sigma}\hat{c}_{m,\sigma},
\end{split}
\label{eq: delta Q operator}
\end{equation}
where the origin of the $x$ axis coincides with the center of the $m=0$ orbital. It measures the charge difference between the left $(x<0)$ and right $(x>0)$ half of the system. 
We compute the expectation value of 
$\hat{Q}_{\mathrm{R}}-\hat{Q}_{\mathrm{L}}$ in the manifold formed by the three lowest states in the $K_y=0$ sector, yielding a $3\times 3$ matrix. Since the product of left-right mirror and time-reversal symmetry leaves the system invariant, the eigenvalues of this matrix are constrained to $\pm q$ and 0, where $q$ is an {\it a priori} unspecified real number. 

\begin{figure}[t]
\begin{center}
\includegraphics[width=0.45\textwidth]{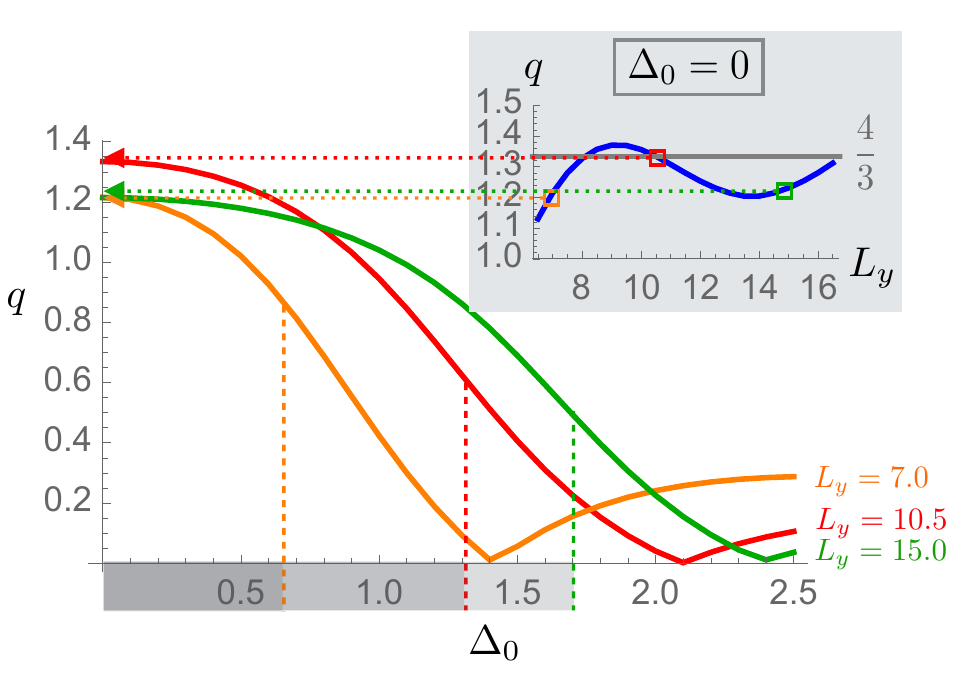}
\caption{
Largest eigenvalue $q$ of the operator $\hat{Q}_{\mathrm{L}}-\hat{Q}_{\mathrm{R}}$, defined in Eq.~\eqref{eq: delta Q operator}, that measures the charge imbalance between the left half and the right half of the cylinder depicted in Fig.~\ref{fig:CylinderOrbitals}~b)
in real space, computed in the manifold of the three lowest energy states with momentum $K_y=0$. 
The data was obtained with Hamiltonian Eq.~\eqref{eq: general Hamiltonian} using the same parameters as in Fig.~\ref{fig:spectra}, except for $L_y/\ell_B=7.0$ (red, $2\pi N_{\mathrm{orb, n}} \ell_B / L_y \simeq 5.4$), $L_y/\ell_B=10.5$ (orange, $2\pi N_{\mathrm{orb, n}} \ell_B / L_y \simeq 3.6$) and $L_y/\ell_B=15.0$ (green, $2\pi N_{\mathrm{orb, n}} \ell_B / L_y \simeq 2.5$) as indicated. Right of the respective dashed lines, the spread-to-gap ratio shown in Fig.~\ref{fig:phase diagrams}~b) is less than one. Moreover the charge imbalance nearly vanishes in the parameter regime of optimal spread-to-spread ratio. The inset shows $q$ as a function of $L_y$ for $\Delta_0=0$.} 
\label{fig:chargeimbalance}
\end{center}
\end{figure}

Figure~\ref{fig:chargeimbalance} shows the evolution of $q$ with the strength of the superconducting pairing $\Delta_0$. For $\Delta_0=0$, we can understand the nearly quantized value $q\approx 4/3$ by recalling that the three lowest $K_y=0$ states are comprised of one state for which both up- and down-spin droplets are centered around $m=0$ and a pair of states in which they are both centered around $m=\pm1$. 
The center of charge of the former is located exactly at $x = 0$, while the latter two states have an excess of charge $\pm q/4$ right of $x = 0$ (i.e., $x>0$) in each layer, and the opposite deficit left of $x=0$ (i.e., $x<0$).
The operator $\hat{Q}_{\mathrm{R}}-\hat{Q}_{\mathrm{L}}$ is thus diagonal in this basis, with respective eigenvalues $0, \pm q$.
In the thermodynamic limit, the $x>0$ excess charge in the latter single layer states is equal to the quasiparticle-excitation charge $\pm 1/3$, summing up to an expectation value $q= 4/3$ of $\hat{Q}_{\mathrm{R}}-\hat{Q}_{\mathrm{L}}$. 
In the absence of superconductivity, the relevant length scales are the perimeter $L_y$ and the width of the droplet $2\pi N_{\mathrm{orb}} \ell_B^2 / L_y$ (where $N_{\mathrm{orb}} = 3\tilde{N}/2 - 2$).
The charge quantization will not be clearly observed unless both these length scales are large compared to the correlation length of the ground state (around $1.4\ell_B$ for the $\nu = 1/3$ Laughlin state~\cite{Estienne-PhysRevLett.114.186801}). In the inset of Fig.~\ref{fig:chargeimbalance}, we indeed observed a nearly quantized $q \approx 4/3$ when tuning the value of $L_y$ to respect this criterion (obtained for $L_y/\ell_B = 10.5$ and $2\pi N_{\mathrm{orb}} \ell_B / L_y \simeq 9.6$).

Upon introducing the superconducting term $\Delta_0 \neq 0$, we observe a rapid decrease in $q$, reaching (and passing) zero near the value of $\Delta_0$ that leads to an optimal spread-to-gap ratio of the nearly degenerate ground states. This can be observed by comparing Fig.~\ref{fig:phase diagrams}~b) and Fig.~\ref{fig:phase diagrams}~c). In the latter, we present $q$ as a function of $\Delta_0$ and the physical distance between the superconducting edges. Two trends can be observed: (i) In the lower right corner of this parameter space, where the edges are in closest spatial proximity, $q$ is the smallest. (ii) In contrast, the largest values of $q$ are found in the upper left corner of this parameter space. However, in this limit of small $L_y$, corresponding to a thin cylinder, the charge distribution strongly varies with $x$ even in the center of the droplet. This yields contributions to the expectation value of $\hat{Q}_{\mathrm{R}}-\hat{Q}_{\mathrm{L}}$ from the center of the droplet $x=0$, such that the operator does not allow for measurement of edge properties only. 

Given the limited system sizes we can study numerically, it is hard to infer the behavior of the system in the thermodynamic limit from this computation. We do, however, present data for one other system size in Supplementary Note I which shares the qualitative features discussed above with Fig.~\ref{fig:phase diagrams}. In summary, we observed that the charge imbalance of the three lowest states in the $K_y=0$ sector evolves from being nearly quantized to $4e/3$ in the limit of two decoupled layers to very small values when the states are nearly degenerate and separated by a gap from other excited states.

\subsection{Spin-dependent flux insertion}
\label{sec: spin-dependent flux}

\begin{figure}[t]
\begin{center}
\includegraphics[width=0.45\textwidth]{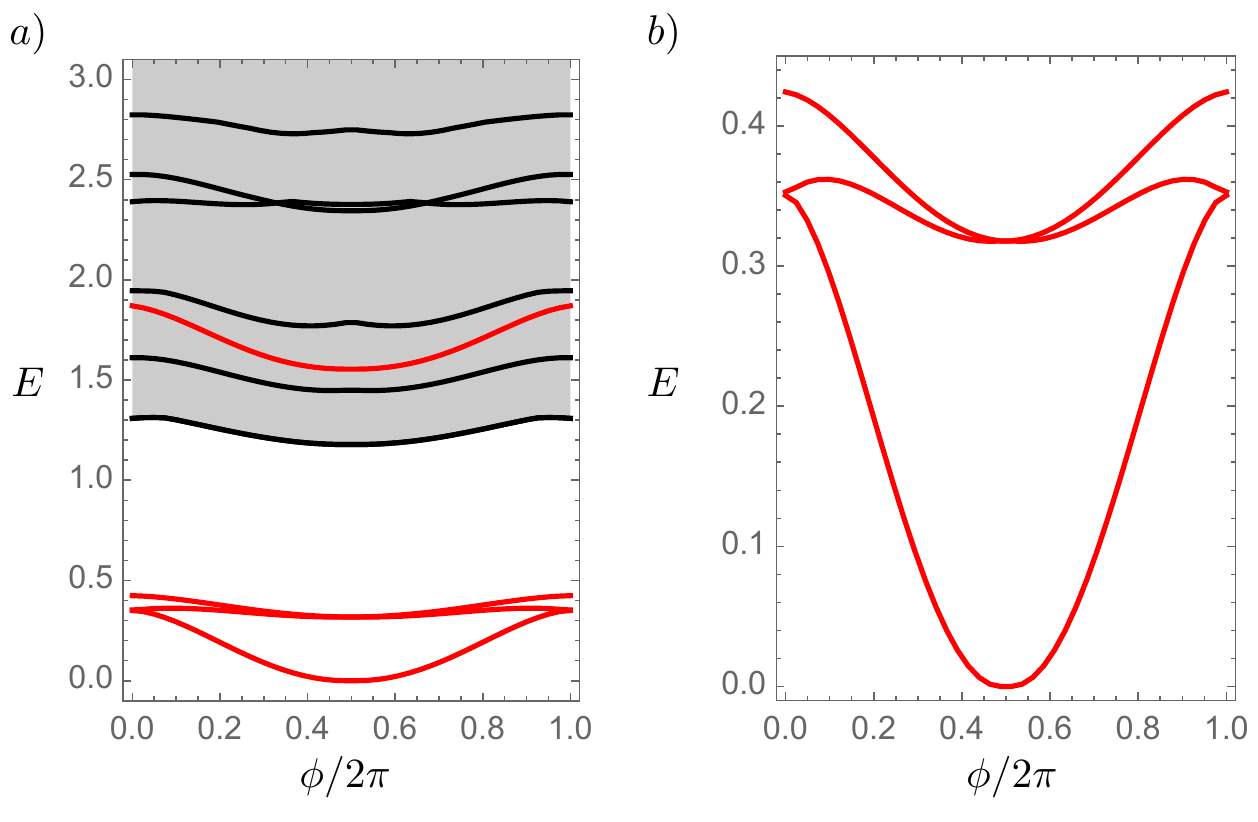}
\caption{
Evolution of the energy levels under spin-dependent flux insertion for Hamiltonian Eq.~\eqref{eq: general Hamiltonian} with the same parameters as in Fig.~\ref{fig:spectra} except for $L_y=7.0/\ell_B$ and $\Delta_0=1.2$.
The spin-dependent flux insertion moves particles in the background of the linear onsite potential, giving rise to the overall $\phi$-dependent energy shift of the eigenstates. 
a) Evolution of the low energy spectrum. Red are the four lowest states in the $K_y=0$ sector, black are the lowest states in each of the other $K_y=0$ sectors. Thus, not all states in the gray region are shown.
b) Close-up of the evolution of the three lowest states corresponding to the topological edge degrees of freedom, showing how the three states are permuted (up to small anticrossings) as $\phi$ is changed by $2\pi$. 
 } 
\label{fig:spinflux}
\end{center}
\end{figure}

To confirm the topological nature of the observed degenerate ground states, we perform a numerical charge pumping experiment. The adiabatic insertion of a magnetic flux $\phi$ along the cylinder axis is equivalent to changing the boundary conditions of the electronic wave functions from periodic to twisted by an angle $2\pi \phi/\phi_0$, where $\phi_0$ is the flux quantum. In a Landau level, as $\phi$ is increased from 0 to $\phi_0$, all single-particle orbitals are shifted by one unit of the quantum number $m\to m+1$. To see this, notice that changing the boundary conditions from periodic to twisted amounts to replacing $m$ by $m+\phi/\phi_0$ in Eq.~\eqref{CylinderObitals}. In a Laughlin 1/3 state, every orbital has an average occupation 1/3, so that, in the thermodynamic limit, a fractional charge $e/3$ is pumped from one end of the cylinder to the opposite end in the process of the flux insertion. 

We would like to utilize a flux insertion to transform the topological ground states, labeled as $|\Psi_{0}\rangle$, $|\Psi_{+}\rangle$, $|\Psi_{-}\rangle$ by their charge imbalance $0$ and $\pm q$, into one another. As we will demonstrate, we can use charge pumping to permute these ground states. Since the two layers of our system are time-reversed partners with opposite Hall conductivities, we have to insert flux with opposite orientation for the $\uparrow$-spin and $\downarrow$-spin particles. Only then is a net charge pumped from one edge of the system to the other. We will refer to this as \emph{spin-dependent flux} insertion [see Fig.~\ref{fig:CylinderOrbitals}~b)]. 

Suppose we start with a state $|\Psi_{0}\rangle$ that has charge 0 on both edges. As unit $\phi_0$ spin-dependent flux is adiabatically inserted, charge is transferred from the left to the right edge, so that the resulting state is $|\Psi_{+}\rangle$. The other ground states are expected to transform into one another analogously: $|\Psi_{+}\rangle\to |\Psi_{-}\rangle$, 
$|\Psi_{-}\rangle\to |\Psi_{0}\rangle$. Thus, after insertion of a quantum of spin-dependent flux, we expect to obtain a permutation of the three ground states. 
This expectation is independent of the presence of a quantization of $q$, because the spectrum has to be invariant under  $\phi\to\phi+\phi_0$. The observation of the state permutation under spin-dependent flux insertion could, however, be obstructed by large avoided crossings in the evolution of the energy levels.
It is important to stress that the spectrum remains gapped (above the three ground states) during the entire process of spin-dependent flux insertion. This gap is provided by the superconducting order parameter that couples states of different particle number on each edge. Without the superconductivity, charge pumping would still occur between the gapless edges, but the adiabatic process would simply accumulate charge at one edge and deplete the other, mapping the eigenstates to others with ever higher energy with each quantum of spin-dependent flux inserted.

To implement the spin-dependent flux insertion, we observe that the substitution $m$ to $m+\phi/\phi_0$ in Eq.~\eqref{CylinderObitals} is equivalent to substituting $m$ by $m+\phi/\phi_0$ in $\mu_m$ and $f_m$ for an infinitely long cylinder. In a finite cylinder, such an approach is still valid for the low-energy subspace as long as the number of orbitals is larger than the number of orbitals typically covered by the incompressible liquid. In our case, this is roughly given by the number of orbitals needed by a single Laughlin state with $\tilde{N} /2$ i.e. $3 \tilde{N} / 2 - 2$.

\begin{subequations}\label{eq: interpolation}
To simulate spin-dependent flux insertion in a system with $N_\Phi+1$ orbitals, $m=-N_\Phi/2,\cdots, N_\Phi/2$, we consider a system enlarged by one orbital, $m=-N_\Phi/2,\cdots, N_\Phi/2+1$ and use a linear interpolation of the functions $\mu_m$ and $f_m$, which allows their argument to take real values and substitute in the Hamiltonian~\eqref{eq: general Hamiltonian}
\begin{equation}
\mu_m\to \mu_{m+\phi/\phi_0}
\end{equation}
and 
\begin{equation}
f_m\to f_{m+\phi/\phi_0}.
\end{equation}
When tuning $\phi$, the potential experiences a kink around $m=0$ which would result in a kink in the energy spectrum. We have thus replaced the absolute value around $0$ by a quartic polynomial interpolation that ensures the potential and its derivative are continuous. Similarly for $f$, we use a linear interpolation  for any orbital at the boundary between a superconducting ($f=1$) and a normal ($f=0$) region. 
\end{subequations}

The low-energy spectrum of the resulting $\phi$-dependent Hamiltonian is plotted in Fig.~\ref{fig:spinflux}. Up to a small avoidance, the three ground states permute as anticipated, while the spectral gap above them stays intact in the process. The overall evolution of all energy levels with a minimum at $\phi=\phi_0/2$ is a result of the specific interpolation~\eqref{eq: interpolation}. Indeed, leaving unoccupied the orbitals near the system ends (where the confining potential is more important) results in a lower total energy. This demonstrates that the superconducting coupling has indeed joined up the two layers into a topological equivalent of the torus geometry sketched in Fig.~\ref{fig:CylinderOrbitals}~c). In Supplementary Note I, we present a full phase diagram for the spin-dependent flux insertion (similar to Fig.~\ref{fig:phase diagrams}) for both this system size and a slightly smaller one.

\subsection{$6\pi$ Josephson effect}
\label{sec: Josephson}

\begin{figure}[t]
\begin{center}
\includegraphics[width=0.45\textwidth]{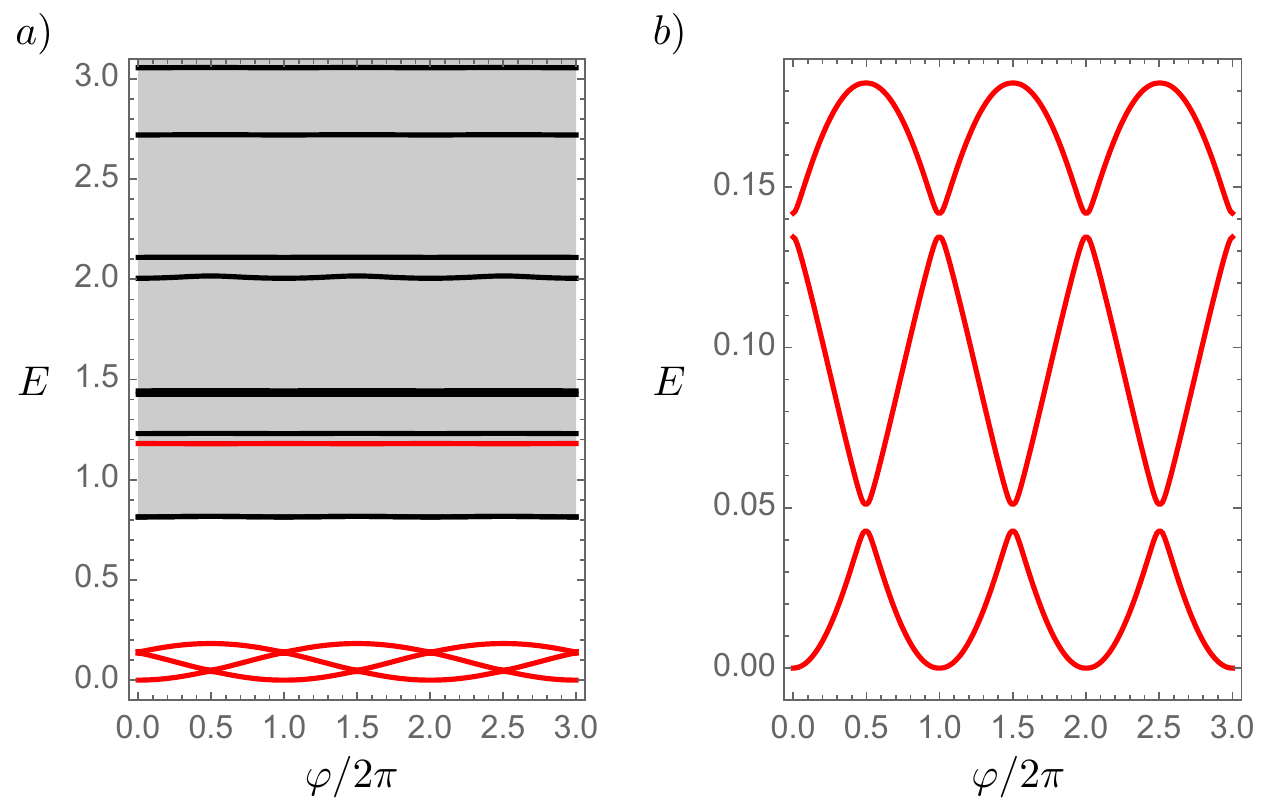}
\caption{
Evolution of the energy levels of the $6\pi$ Josephson effect for Hamiltonian Eq.~\eqref{eq: general Hamiltonian} with the same parameters as in Fig.~\ref{fig:spectra} and $\Delta_0=2.13$, varying the phase difference $\varphi$ between the left and the right superconducting edge.  
a) Evolution of the low energy spectrum. Red are the four lowest states in the $K_y=0$ sector, black are the lowest states in each of the other $K_y=0$ sectors. Thus, not all states in the gray region are shown.
b) Close-up of the evolution of the three lowest states corresponding to the  topological edge degrees of freedom, showing a $6\pi$ periodicity with a small avoidance of the crossings between the states. 
 } 
\label{fig:Jospehson}
\end{center}
\end{figure}

As a second piece of evidence that the heterostructure realizes the  topological superconducting edges, we calculate the evolution of the energy levels that corresponds to the $6\pi$ Josephson effect. In order to do so, the relative complex phase between the mean-field superconducting order parameters on the left and the right edge, $\varphi$, is varied. 
The Josephson effect requires quasiparticle tunneling processes between the superconducting edges.
Necessarily, the \emph{spectrum} of the heterostructure displays a $2\pi$ periodicity in $\varphi$. However, 
 in the thermodynamic limit, in which the three states are degenerate, the \emph{ground state} of the system does not return to itself when $\varphi$ is advanced by $2\pi$. Rather, it evolves into another degenerate ground state and only after $\varphi$ is advanced by $6\pi$ does the system return to its initial state.
The reason for this behavior is that the elementary excitations of the superconducting edge are Cooper paired quasiparticles of charge $2e/3$, delocalized along the cylinder perimeter, which tunnel across the bulk gap.

To observe the $6\pi$ Josephson effect numerically in our finite-size setup, the energy scale associated with the tunneling must be larger than the finite size splitting between the ground states. For the system sizes accessible to exact diagonalization calculations, this is not generically the case, even when the spread-to-gap ratio shown in Fig.~\ref{fig:phase diagrams}~b) is large. Since the tunneling amplitude is exponentially small in the distance between the edges, we expect a favorable regime for large cylinder circumference $L_y$ (at fixed number of non-superconducting orbitals), so that the physical distance $\propto L_y^{-1}$ between the edges is small. 
Figure~\ref{fig:Jospehson} shows the spectral evolution as a function of $\varphi$ in this regime. We observe that the three low-lying states are indeed permuted as $\varphi$ advances by $2\pi$ up to a residual small avoidance of the crossings between the states. 

To investigate in which region of phase space this type of spectral evolution can be found, we plot the ratio $r$ of largest avoided crossing over energy spread of the ground state manifold, i.e. the quantity
\begin{equation}
r:=
\frac{\mathrm{max}\left[(E_2(0)-E_0(0)), (E_1(\pi)-E_0(\pi))\right]}{E_2(\pi)-E_0(0)},\label{eq:RJosephson}
\end{equation}
where $E_0(\varphi)$, $E_1(\varphi)$, $E_2(\varphi)$ are the energies of the three lowest states as a function of flux $\varphi$. When the avoidance in the evolution of the energy levels vanishes, such that $E_1(0)=E_2(0)$ and $E_0(\pi)=E_1(\pi)$, then $r\to0$ and the $6\pi$ Josephson effect becomes clearly observable. In the opposite limit where the low-energy spectrum is essentially independent of $\varphi$ $E_i(0)=E_i(\pi)$, $i=0,1,2$, we have $r\to 1$ and the tunneling matrix elements are too small to overcome the finite-size induced energy splitting between the three low-lying states.
Figure~\ref{fig:phase diagrams}~d) shows $r$ as a function of the strength of the superconducting pairing potential and the physical distance between the edges. Indeed, we find that the $6\pi$ Josephson effect is best observed when both the spread-to-gap ratio and the distance between the edges is smallest. 

\section{Discussion}

\begin{figure}[t]
\begin{center}
\includegraphics[width=0.37\textwidth]{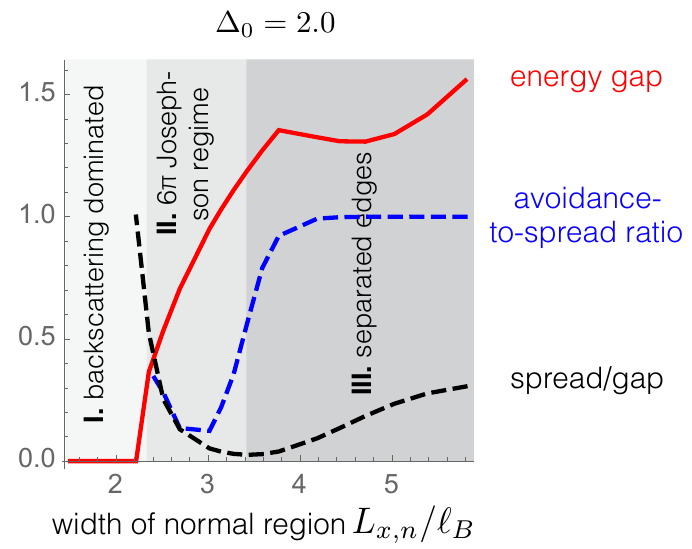}
\caption{Phase diagram along the $\Delta_0=2.0$ line. The horizontal axis is the width of the normal region $L_{x,n}=\frac{2\pi \ell_B^2 N_{\rm orb,n}}{L_y}$. The vertical axis is either the energy gap (red), the spread over gap ratio (dashed black), or the avoidance-to-spread ratio for the Josephson effect (dashed blue). We clearly discriminate three different regimes: {\bf I} for small $L_{x,n}$, the backscatteting dominated regime with the breakdown of the gapped edge modes; {\bf II} the intermediate region corresponding to the $6\pi$ Josephson regime; {\bf III} at large $L_{x,n}$, the separated gapped edge modes.}
\label{fig:delta0_2}
\end{center}
\end{figure}

We numerically studied a heterostructure of a bilayer system of FQH Laughlin states with counter-propagating edge modes that are gapped out by a mean-field superconducting order parameter. The system in the cylinder geometry with two gapped edges realizes a nonlocal topological qutrit. 

Our calculations were performed using exact diagonalization and by restricting the computation to the quasihole subspace of the Laughlin state in each layer. Despite this simplification, the system size is still limited. Nevertheless, we have been able to demonstrate four key features : (i) the edges develop a spectral gap induced by the superconducting coupling, (ii) the expected number of three nearly degenerate ground states without any charge imbalance between the two halves of the system, (iii) that charge pumping can permute the ground states, and (iv) that the system exhibits a $6\pi$-periodic Josephson effect. For each signature, we discussed the suitable parameter regime.

While the details of the phase diagram are affected by important finite-size effects, similar features can be identified in all studied systems for $\Delta_0 \gtrsim 1.5$.
Extrapolating from these features, we propose a physical summary and a highlight of our quantitative results on the $6\pi$ Josephson effect in Fig.~\ref{fig:delta0_2}. We focus on a given value of the pairing parameter $\Delta_0=2.0$ and we explore the different regimes as a function of the size of the normal region or equivalently the distance between the two superconducting leads. This distance $L_{x,n}$ is deduced from the number of orbitals without superconducting coupling as discussed in Sec.~\ref{sec:gapping}. As seen in Fig.~\ref{fig:delta0_2}, the value of $L_{x,n}$ determines the behavior of the system among three regimes. When $L_{x,n}\gtrsim 3.4 l_B$, there is a large gap and an approximate threefold degeneracy, leading to well defined gapped edge modes. But the tunneling required for the $6\pi$ Josephson effect is exponentially suppressed. In the intermediate regime $2.4 l_B \lesssim L_{x,n}\lesssim 3.4 l_B$, the $6\pi$ Josephson effect is clear and a robust gap remains. Note that the optimal value of $L_{x,n}$ is roughly twice the correlation length\cite{Estienne-PhysRevLett.114.186801} $\xi \simeq 1.4 l_B$ of the Laughlin $\nu=1/3$ phase which allows tunneling without destroying the underlying quantum liquid. Finally, when $L_{x,n}$ is small, i.e., a distance lower than the correlation length $\xi$, the induced gap at the edges collapses. Thus our results validate the hypothesis of the previous effective approaches and provide an estimate of the characteristic dimensions for future experimental implementations.

Our work provides the first quantitative study of fractional edge modes coupled to superconducting leads in a fully microscopic model. Future works, most probably relying on the density matrix renormalization group calculations, should be able to rely on our setup and results to provide new insights. In particular, it could overcome the size limitation and address the potential new emerging phases when substituting the Laughlin state with any richer topological order.

\begin{acknowledgments}
We are very grateful to Maissam Barkeshli for several inspiring discussions during the course of this work and for helpful comments on the manuscript.
We thank C.~Nayak, B.~A.~Bernevig, C.~Mora and B.~Estienne for fruitful discussions. C.~R. was partially supported by the Marie Curie programme under ED Grant agreement 751859. N.~R. was supported by Grant No.~ANR-16-CE30-0025. A.~M.~C. also wishes to thank the Aspen Center for Physics, which is supported by National Science Foundation grant PHY-1066293, and the Kavli Institute for Theoretical Physics, which is supported by the National Science Foundation under Grant No. NSF PHY-1125915, for hosting during some stages of this work.T.~N. acknowledges support from the European Research Council (ERC) under the European Union’s Horizon 2020 research and innovation programm (ERC-StG-Neupert-757867-PARATOP).
\end{acknowledgments}

\section*{Author contributions}
CR and NR developed the numerical code.
CR, AC, and NR performed the numerical calculations.
All authors contributed to the analysis of the numerical data and to the writing of the manuscript.

%%\newpage

%%\newpage

\newpage

\begin{center}
{\bf Supplementary Material}
\end{center}

\appendix

\section{Results for a different system size}\label{app:NumericalResults}

\begin{figure*}[t]
\begin{center}
\includegraphics[width=0.98\textwidth]{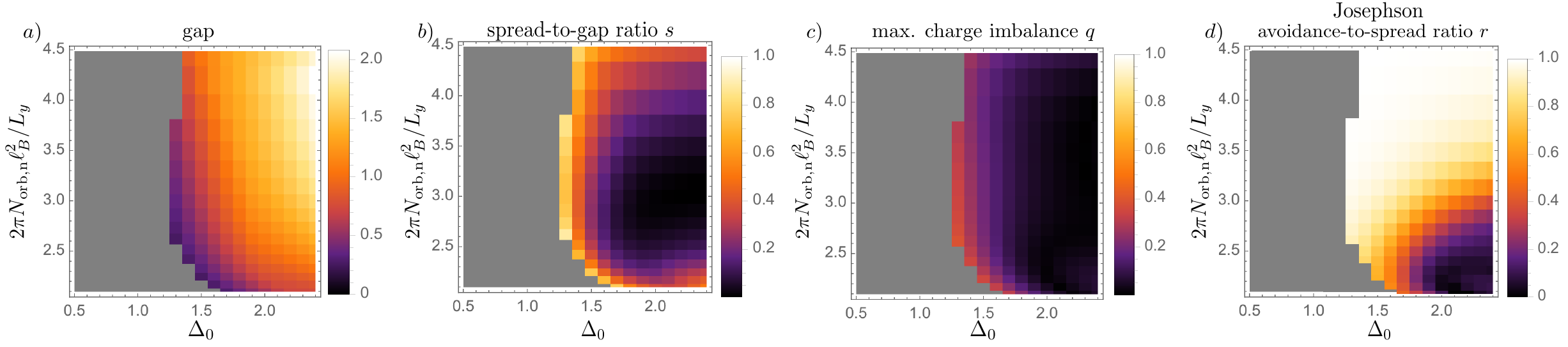}
\caption{
Characterization of the low-energy spectrum of Hamiltonian Eq.~\eqref{eq: general Hamiltonian} as a function of $2\pi N_{\mathrm{orb,n}}\ell_B^2/L_y$, which is the physical distance between the superconducting regions and changed by varying $L_y$, 
and the strength of the pairing potential $\Delta_0$, with parameters $N_\phi=20$ and $N_{\mathrm{orb,sc}}=16$, as well as a mean number of particles $\bar{N}=10$.
a) Gap between the three lowest states in the $K_y=0$ sector and the next excited states. Gray color indicates that the three lowest states do not have $K_y=0$.
b) Spread of the three lowest states in the $K_y=0$ sector as indicated in Fig.~\ref{fig:spectra}~c) divided by the gap. Gray color indicates the region in which the ratio exceeds 1 or where the three lowest states do not have $K_y=0$.
c) Maximal eigenvalue of the  operator measuring the charge imbalance between the left and right half of the system in the space of the tree lowest states. Gray color is used if the three lowest states do not have $K_y=0$.
d) The difference between the energy of the second lowest eigenstate at Josephson phases $\varphi=0$ and $\varphi=\pi$, normalized by the spread and each time measured with respect to the ground state energy. The closer this quantity gets to  1, the better can the $6\pi$ Josephson effect be observed. The gray color has the same meaning as in c)
 } 
 \label{fig:phase diagrams different size}
\end{center}
\end{figure*}

In the article, we have focused on a single system. This system was the largest one for which a complete study could be performed. We have considered system sizes, with different numbers of superconducting orbitals and various number of particles $\bar{N}$. Here we provide additional numerical results for a smaller system size of $N_\phi=20$ in Fig.~\ref{fig:phase diagrams different size}. The charging energy parameters are optimized such that the ground state in the absence of superconducting coupling has now a number of particles $\bar{N}=10$. When compared to Fig.~\ref{fig:phase diagrams} for $N_\Phi=21$ and $\bar{N}=12$, this smaller system has proportionally more superconducting orbitals ($N_{\mathrm{orb,sc}}=16$) while preserving an equivalently normal region size. Note that a similar system for $\bar{N}=12$ would require $N_\phi=23$ and $N_{\mathrm{orb,sc}}=18$ which is unfortunately numerically out of reach at the moment.

As a consequence of having more superconducting orbitals, we now observe a much larger region where the $6\pi$ Josephson effect can be observed [Fig.~\ref{fig:phase diagrams different size}~d)]. It still requires the two physical edges to be close but it appears for a wider range of superconducting coupling strength. Having a smaller system is twofold has nevertheless some drawback. Indeed, the gap (Fig.~\ref{fig:phase diagrams different size}~a) and the spread over gap ratio (Fig.~\ref{fig:phase diagrams different size}~b) have similarities with the bigger system discussed in the main text. This is especially valid in the range where the normal region has a size lower than $4l_B$. Note that plotting the numerical data as a function of the normal region width helps a lot to compare different system sizes. But the agreement is getting lower when we increase the normal region size, with a much weaker gap and a worse $s$. 

\begin{figure}[t]
\begin{center}
\includegraphics[width=0.48\textwidth]{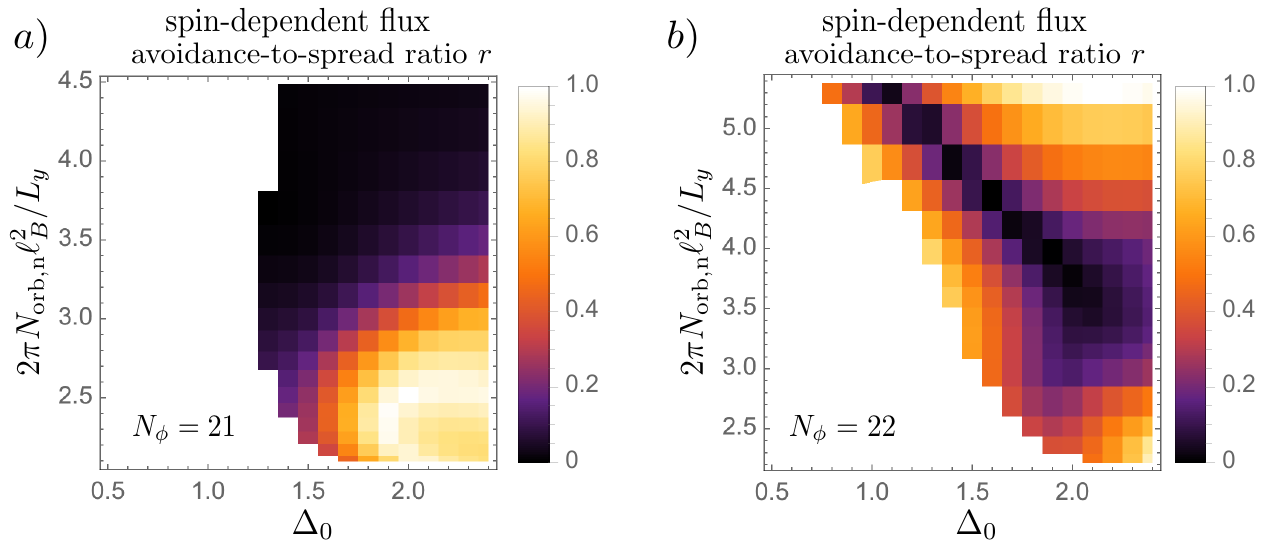}
\caption{
Avoidance-to-spread ratio $r$ for spin-dependent flux, defined in Sec.~\ref{app:NumericalResults}, for Hamiltonian Eq.~\eqref{eq: general Hamiltonian} as a function of $2\pi N_{\mathrm{orb,n}}\ell_B^2/L_y$, which is the physical distance between the superconducting regions and changed by varying $L_y$, 
and the strength of the pairing potential $\Delta_0$, a) with parameters $N_\phi=21$ and $N_{\mathrm{orb,sc}}=17$, as well as a mean number of particles $\bar{N}=10$ b) with parameters $N_\phi=22$, $N_{\mathrm{orb,sc}}=17$ and a mean number of particles $\bar{N}=12$.
Gray color is used if the three lowest states do not have $K_y=0$. The dark diagonal line in b) is an artifact of the spread being extremely small along this line.
 } 
\label{fig:spin-dependent flux quality}
\end{center}
\end{figure}

We also provide a phase diagram for the spin-dependent flux insertion in Fig.~\ref{fig:spin-dependent flux quality} both for the system size addressed here [Fig.~\ref{fig:spin-dependent flux quality}a)] and for the system size discussed in the main text [Fig.~\ref{fig:spin-dependent flux quality}b)]. We again use the ratio $r$ (defined in Eq.~\eqref{eq:RJosephson}) with flux $\varphi$ replaced by spin-dependent flux $\phi$. A small value of $r$ indicates the regime where the spin-dependent flux insertion is clearly observable.
As can be seen in Fig.~\ref{fig:spin-dependent flux quality}, we find a large region that does not even require the two edges to be too far from each other (albeit not too close) and where there is a clear mixing of the three states under the spin-dependent flux insertion.

\section{Parameter determination for the charging energy}\label{app:ChargingEnergy}

Here we explain how we determine the parameters entering the charging energy term, $C$ and $N_0$.
We demand that the ground state is in the sector with some target number of particles $\bar{N}$ and that the lowest energy eigenstates of Hamiltonian~\eqref{eq: general Hamiltonian} for $\Delta_0=0$ in the particle number sectors $\bar{N}-2$ and $\bar{N}+2$ are degenerate. Denoting the energy of the chemical potential term of the Hamiltonian as $E^{(0)}_{N}$ for $N$ particles, this yields the condition
\begin{equation}
E^{(0)}_{\bar{N}+2}+\frac{1}{2C}(\bar{N}+2-N_0)^2
=
E^{(0)}_{\bar{N}-2}+\frac{1}{2C}(\bar{N}-2-N_0)^2.
\end{equation}
We solve this to determine the charging energy as
\begin{equation}
\frac{1}{2C}=\frac{E^{(0)}_{\bar{N}+2}-E^{(0)}_{\bar{N}-2}}{8(N_0-\bar{N})}.
\end{equation}
The gap between the lowest state in the $\bar{N}$ particle sector and the degenerate lowest states in the 
$\bar{N}\pm2$ particle sectors is then given by
\begin{equation}
\Delta_{\pm2}=\frac{E^{(0)}_{\bar{N}+2}-2 E^{(0)}_{\bar{N}}-E^{(0)}_{\bar{N}-2}}{2}
+\frac{E^{(0)}_{\bar{N}+2}-E^{(0)}_{\bar{N}-2}}{N_0-\bar{N}}.
\end{equation}
The first term is the discrete second derivative of the potential energy term, which is bounded from below by the finite size quantization of the edge energy levels. The second term, a charging energy contribution can be minimized by choosing $N_0$ large. Thus, a large $N_0$ makes the particle number fluctuation modes in the system softer. 
 
\bibliography{MS_EDparafermions.bib}

\end{document}